\documentclass[12pt,preprint]{aastex}

\slugcomment{}

\shorttitle{On the Detection of Two New Transneptunian Binaries from the CFEPS Kuiper Belt Survey. }
\shortauthors{Lin et al.}

\begin{document}

\title{On the Detection of Two New Transneptunian Binaries from the CFEPS Kuiper Belt Survey. }

\author{H.-W. Lin\altaffilmark{1}}
\affil{Institute of Astronomy, National Central University, Taiwan}
\email{sevenlin123@gmail.com}

\author{J.J. Kavelaars\altaffilmark{2}}
\affil{Herzberg Institute for Astrophysics, Canada}

\author{W.-H. Ip\altaffilmark{1}}
\affil{Institute of Astronomy, National Central University, Taiwan}

\author{B.J. Gladman\altaffilmark{3}}
\affil{Dept. of Physics and Astronomy, University of British Columbia, Canada}

\author{J.M. Petit\altaffilmark{2,4}}
\affil{Dept. of Physics and Astronomy, University of British Columbia, Canada}\affil{Observatoire de Besancon, France}

\author{R. L. Jones\altaffilmark{2,3}}
\affil{Herzberg Institute for Astrophysics, Canada}\affil{Dept. of Physics and Astronomy, University of British Columbia, Canada}

\and

\author{Joel Wm. Parker\altaffilmark{6}}
\affil{Space Science \& Engineering Division, Southwest Research Institute, USA}

\altaffiltext{1}{Institute of Astronomy, National Central University, Taiwan}
\altaffiltext{2}{Herzberg Institute for Astrophysics, 5071 West Saanich Road Victoria, BC V9E 2E7, Canada}
\altaffiltext{3}{Department of Physics and Astronomy, 6224 Agricultural Road, University of British Columbia, Vancouver, BC, Canada}
\altaffiltext{4}{Observatoire de Besanon, B.P. 1615, 25010 Besanon Cedex, France}
\altaffiltext{5}{Space Science \& Engineering Division, Southwest Research Institute, 1050 Walnut Street, Suite 400, Boulder, CO 80302, USA}

\begin{abstract}

We report here the discovery of an new near-equal mass Trans-Neptunian Binaries (TNBs) L5c02 and the and the putative detection of a second TNB (L4k12)among the year two and three detections of the Canada-France-Eclipic Plane Survey (CFEPS).
These new binaries (internal designation L4k12 and L5c02) have moderate separations of $0.4\arcsec$ and $0.6\arcsec$ respectively. The follow-up observation confirmed the binarity of L5c02, but L4k12  are still lack of more followup observations.
L4k12 has a heliocentric orbital inclination of $\sim 35^{\circ}$, marking this system as having the highest heliocentric orbital inclination among known near-equal mass binaries. 
Both systems are members of the classical main Kuiper belt population.  

Based on the sample of objects searched  we determine that the fraction of near-equal mass wide binaries with separations $> 0.4\arcsec$ is $1.5\%$ to $20\%$ in the cold main classical Kuiper belt and, if our detection the binarity L4k12 holds, $3\%$ to $43\%$ in the hot main classical objects are binary. 
In this manuscript we describe our detection process, the sample of objects surveyed, our confirmation observations.

\end{abstract}


\section{Introduction}

Just as the detection of the first trans-Neptunian object (TNO, neglecting Pluto), 1992 QB$_{1}$, by \citet{Jew92} let to an avalanche of new TNO discoveries, the serendipitous discovery of the first trans-Neptunian binary (TNB, neglecting the Pluto-Charon system), 1998 WW$_{31}$, by \citet{Vei02} spurred the discovery of dozens more such systems.  There are currently more than 50 TNBs known \citep{Nol08a}.  The growing catalogue of known TNOs and TNBs discovered in characterized surveys permits the use of these objects as probes of structure formation and evolution in the outer solar system.

The large sky separation, $1.2^{\arcsec}$ and the similar brightness, $\Delta_{mag} \sim 0.4$, of the two components of the 1998 WW$_{31}$ provided the first clue to the enigmatic nature of TNBs.
That this was the first kind of TNB discovered, however, is partially due to the observational requirement of being able to split the two sources in-order to determine that one is observing a binary. 
Since the TNOs are rather faint, most observations are of rather low SNR ($\sim 10$) and to detect a TNB in such observations generally requires that the two components have an angular separation of order the seeing of the observations.  
The next TNB discovered, 2001 QT$_{297}$ \citep{Osi08} also had large separation that was at the limit of the resolution of the observations. 
The discovery separation of 2001 QW$_{322}$ \citep[$\sim 4\arcsec$,][]{Pet08} was much larger than the seeing of the observations, perhaps indicating that large separation binaries are more than just an artifact of the observing circumstances.  
At this time about $\sim 10\%$ of the known TNBs have separation in excess of $1\arcsec$.  The
 pathway of wide separation TNBs are, as yet, not well understood.
In general, the study of binary systems provides very direct information on the physical nature of the constituent members. 
Binary systems provide the only reliable means to derive the masses of TNOs. 
Furthermore, if their sizes can be computed by using albedos obtained from optical and infrared measurements, their average bulk densities can be estimated to to roughly $10\%$ accuracy. 
While these measurements provide data on the physical characteristics of the binary components, knowledge of the physical properties for the TNO population provide inputs to understanding the formations of the TNB themselves.

In the cold classical disk of TNOs inclinations $< 5^{\circ}$, the occurrence rate of TNBs with separation larger than 0.1\arcsec is about 17-32\% while only 3-10\% for the other dynamical classes \citep{Ste06, Nol08a, Nol08b}. 
The differences in the proportion of TNBs could be the result of the dynamical histories of different TNO populations, indicative of different formation histories or survival rates for TNBs in the two populations.
To help understand the difference in TNB properties and thus unravel the formation conditions of the outer solar system more attention should must be given to the census of different types of TNBs (i.e., wide binaries vs. close binaries) in different regions of the Trans-Neptunian belt. 

Several large-scale ground-based surveys of TNOs have been performed or in process. These include the Deep Ecliptic Survey (DES) at CTIO and Magellan \citep{Mil02, Ell05}, the Keck Telescope survey \citep{Tru03}, and the Canada-France Ecliptic Plane Survey (CFEPS, \citep{Jon06, Kav09, Pet10}). The first two surveys have yielded three TNBs out to a total of 362 TNOs \citep{Nol08a}. In this work we will report the putatitve detection of two new TNBs based on the CFEPS observations. The observations used in our TNB search and our search technique are described in Section 2.1 and 2.2. The results of our search and follow-up observations are given in Section 2.3 and 2.4. Section 3 provides a discussion and summary.

\section{Observation and Analysis}
\paragraph{}

The Canada-France Ecliptic Plane Survey (CFEPS) used the observations of the Very Wide sub-component of the Canada-France-Hawaii Telescope Legacy Survey (CFHT-LS).  
The CFEPS Kuiper belt study is a primary scientific goal of the CFHLS-VW (Legacy Survey- Very Wide) which employs the MegaCam camera with a $1^{o} \times 1^{o}$ field of view. 
The purpose is to provide a Kuiper Belt survey with quantitatively-known detection
biases and long-term tracking to achieve precise orbits of KBOs. 
CFEPS has surveyed fields whose ecliptic latitudes range between $+/- 2^{o}$ from the ecliptic plane. 
Approximately 500 square degrees have been covered with a detection limit of about 24.3 magnitudes in g' and subsequently re-imaged, for tracking, in r' and i'.   At completion of the CFHT-LS  component of the survey approximately $200$ TNOs had been discovered and tracked to three oppositions, here we report on the search for TNBs among these TNOs \citep{Pet10}. 

The CFEPS images of newly discovered TNOs can be used to determine if any of the TNOs are in fact TNBs. The most direct method is to look for separate components of the binary systems, as has been done by previous authors. A visual search of the discovery images for TNBs is only sensitive to large ($\gtrsim 1\arcsec$) separation TNBs, because of the limitation in angular resolution of the detector ($\sim 0.187\arcsec$ per pixel) and seeing condition ($\gtrsim 0.7\arcsec$).  No TNBs were found during the initial visual inspection of the CFEPS discovery images.  

\subsection{Point Source Modeling}

\begin{figure}
\plotone{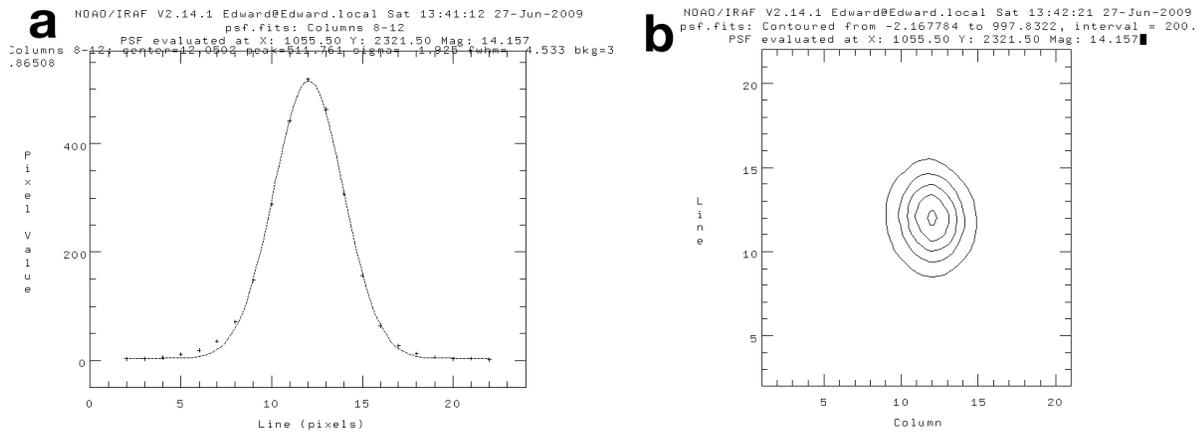}
\caption{An example of the PSF of single stars in a CFEPS field which is constructed by averaging the PSFs of several single stars around a TNO: (a) a cross-section of the PSF; (b) a two-dimensional projection. }
\label{fig:psf}
\end{figure}

To improve our sensitivity to smaller separation TNBs ($\lesssim 1 \arcsec$) we examine and compare the point-spread functions of individual KBO images to check if they are truly point sources or possible extended sources similar to that employed in previous TNB searches \citep{Ker06, Ste06}.
The IRAF\footnote{IRAF is distributed by the National Optical Astronomy Observatory, which is operated by the Association of Universities for Research in Astronomy (AURA) under cooperative agreement with the National Science Foundation.} DAOPHOT package was used to build a point-source model (PSF) for each of the three CFHT/MegaCam images in which a CFEPS TNO was discovered.  
Several (10-15) isolated bright(SNR $\gtrsim 50$) point-sources were hand selected as the model inputs.  
The DAOPHOT package computes the average shape of the point sources to obtain the PSF model (Figure~\ref{fig:psf}~).  
The model input stars were selected to be within $\sim 1\arcmin$ from the target TNO to avoid the distortion across the the CFHT/MegaCam field. 

The utility of the PSF modeling approach to binary detection can be determined via construction of artificial binaries.  
The IRAF/ADDSTAR task was used to create artificial near equal magnitude TNBs by adding a series of closely separated doublet sources whose total fluxes are similar to those of the TNOs being examined. 
These simulation reveal that TNBs with separations larger than about $80\%$ of the FWHM of the image have goodness-of-fit $\chi^{2}$ values more than 2-$\sigma$ larger than the $\chi^{2}$ for point sources in the same image.  Artificial binaries with separations smaller than 80\% of the FWHM for the image could not be distinguished from point sources using this method.

For each image of a given TNO the $\chi^{2}$ goodness-of-fit parameter was determined. 
Sources whose value of $\chi^{2}$ differed significantly from that of similar brightness point sources in the image were flagged as possible TNBs.
Visual examination of the flagged candidates revealed that in all but one case the observed large residuals were caused by proximity to background sources and not a close companion. 
In one case, L5c02, the deviant $\chi^{2}$ value appeared to be associated with a close companion (see Figure~\ref{fig:compare}).  
The large number of false positives and the low rate of successful detection is likely due to the combination of low SNR of the CFEPS TNO detection images coupled with the low-frequency of large ($> 0.7\arcsec$) separation TNBs.

\subsection{Source elongation}

Examination of the image shape along a single spatial-dimension using a simple functional form, provides an improved approach for binary detection. 
This approach reduces the number of degrees of freedom (as compared to the PSF fitting approach) while increasing the available signal by collapsing the source profile along one spatial dimension. 
A one-dimensional Gaussian provides a reasonable match to astronomical images and has only three degrees of freedom: total flux, centroid, and width. 

Using the ellipse fitting routine in IRAF/ IMEXAM the ellipticity and position angle of each TNO can be determine.  
The flux along the minor axis of the TNO image is then collapsed and the best-fit Gaussian shape parameters along the major axis are determined. 
The width of the Gaussian that best matches the collapsed TNO light profile is then compared to the distribution of widths for point sources in the image, also collapsed along the same direction as the TNO image.
The deviation of the width computed for the TNO from the expected value based on the reference point sources is used to flag possible TNBs.
Those sources with width more then 2-$\sigma$ larger than the mean for point sources in the same frame were examined visually.  
Again, all TNOs flag for visual inspection where found to have background contamination except for two sources: L5c02 (previously detected as binary using PSF modeling) and perhaps L4k12.
Tests with artificially created near equal magnitude TNBs revealed that this one-dimensional fitting approach is sensitive enough to detect binaries with separations larger than 1/3 the FWHM.

\subsection{Detected TNBs}

\begin{figure}
\plotone{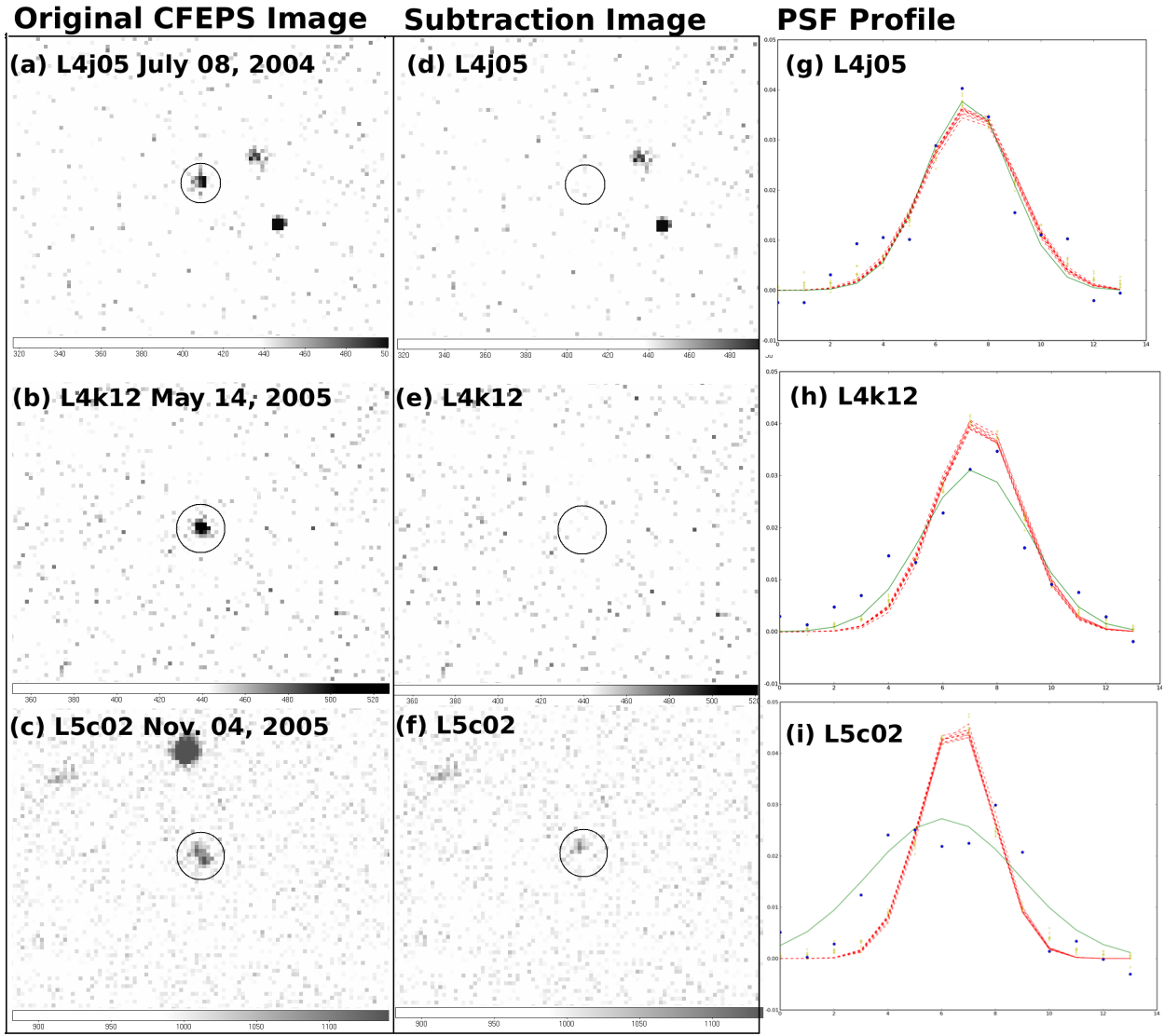}
\caption[Compare]{A comparison of (a-c) the original images of three TNOs L4j05(non-detection), L4k12 and L5c02, which are detected by CFEPS, (d-f) their images after subtraction from the reference PSF, and (g-i) the fitting of their PSF profiles(solid lines) along the elongated axis to the reference PSFs(dash lines). The normalize pixel values of TNOs are shown in circle dots.}
\label{fig:compare}
\end{figure}

Figure~\ref{fig:compare} presents a comparison of the image subtraction and elongation fitting of three CFEPS TNOs: L4j05, L4k12 and L5c02.  
Two of the sources in Figure~\ref{fig:compare} (L4j05, L4k12) have residuals that are normal for point sources in their images and values of $\chi^{2}$ consistent with a point source. 
The TNO L4j05 is found to have an elongation width consistent with point sources detected in the same frame and is presented as an example of a none-detection. 
L4k12, however, exhibits significant deviation from the corresponding one-dimensional Gaussian fitting of point sources in the image, revealing the extended nature of this source.  The TNO L5c02 exhibits obvious residuals compared to the PSF model and also exhibits one-dimensional Gaussian shape which is much broader than the points sources in the same frame.

Using these two procedures (PSF modeling fitting and profile elongation) we identified two binary candidates (L5c02 and L4k12) out of 63 TNOs examined. Those 63 TNOs are the brightest TNOs from CFEPS samples ($r \lesssim 23.0$).
Table~\ref{tab:log} gives the orbital parameters and photometric data of these two objects. 
Based on the classification system of \citet{Ga08}, both L5c02 and L4k12 are members of the classical main Kuiper belt. That is, neither are scattering or have a resonant orbit with Neptune, and they have semimajor axes between the 3:2 and 2:1 mean-motion resonances \citep{Ga08}.

\subsection{Follow-up Observations}

In order to confirm the binary nature of L4k12 and L5c02, we have obtained high-resolution follow-up images using the Subaru 8-meter and WIYN 3.5-meter observatories.

\begin{figure}
\plotone{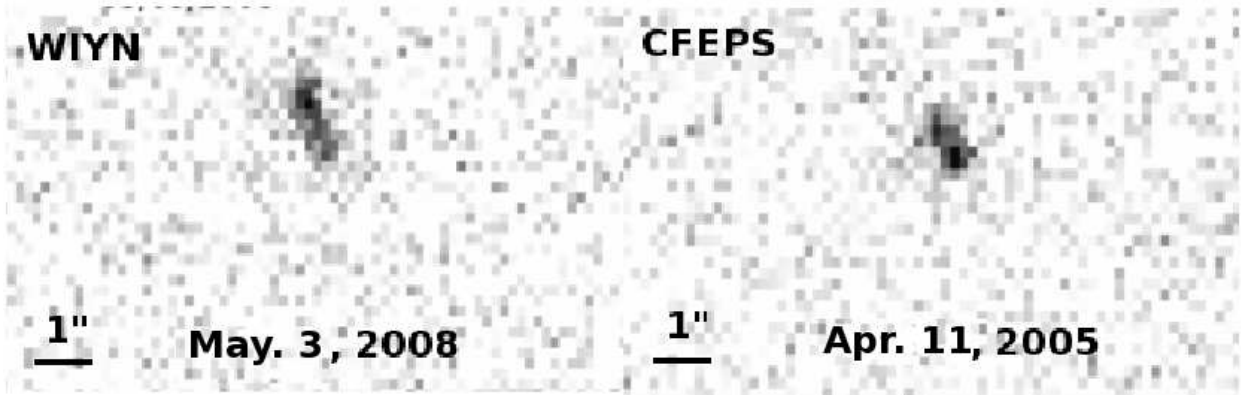}
\caption[L5c02]{The confirmation image of L5c02 taken by WIYN (3.5 m at Kitt Peak) on May 3, 2008. Also shown are the CFEPS discovery image (Apr. 1, 2005) plotted in the same angular size for comparison. }
\label{fig:L5c02}
\end{figure}

Observations of L5c02 obtained with the WIYN telescope on 3 May 2008 are presented in Figure~\ref{fig:L5c02}. 
An elongated structure with a length of about $0.67\arcsec$ (21,000 km at 44.2 AU) is clearly revealed in this image. 
The two components from this image are not fully resolved, however, simultaneous PSF modeling using two points sources reveals that this object is well represented by two sources of comparable magnitudes.

\begin{figure}
\plotone{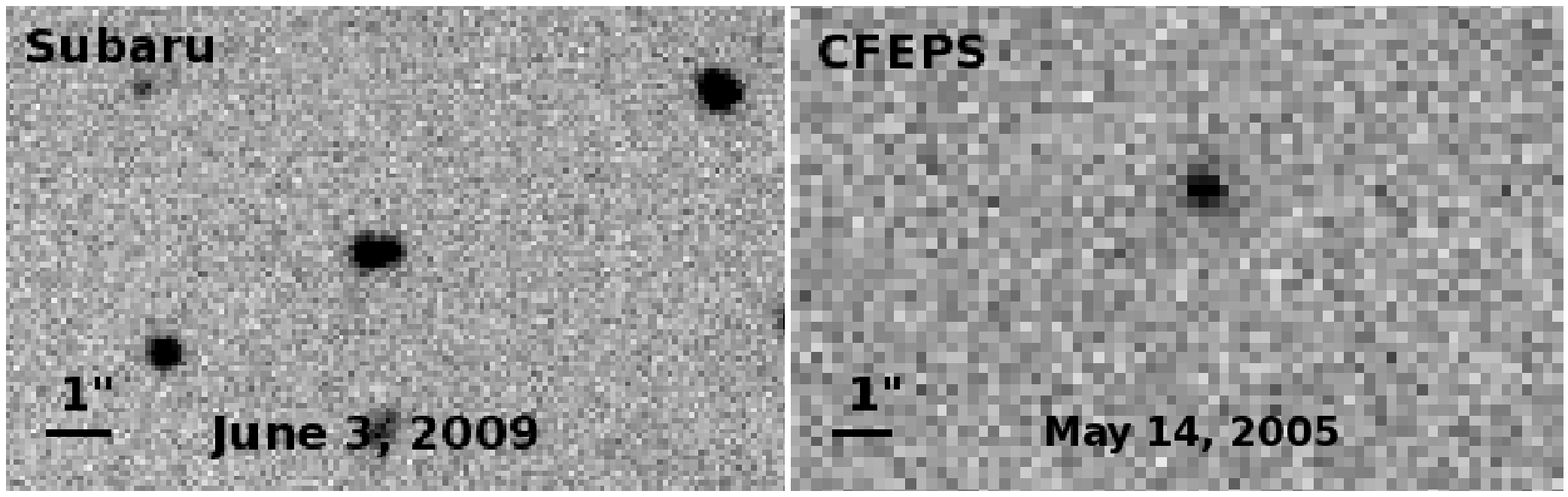}
\caption[L4k12]{The single image of L4k12 taken by Subaru on April 23, 2009. Also shown are the CFEPS discovery image (May 14, 2005) plotted in the same angular size for comparison. }
\label{fig:L4k12}
\end{figure}

Observations of L4k12, obtained with Subaru on April 23, 2009, are shown in Figure~\ref{fig:L4k12}. 
An elongated structure with a major axis of about $0.8\arcsec$ (25,000 km at 43.6 AU) is quite prominent in the images. Although the $0.4\arcsec$ to $0.8\arcsec$ separation change is plausible, it is possible this single Subaru image is of a background star behind the TNO.  If the elongated structure is not due to background star behind the TNO, this image is well modeled by doublet source with both sources having comparable magnitudes. 

Table~\ref{tab:follow} provides a summary of the information on the CFEPS discovery, recovery and follow-up observations. 

\section{Discussion and Summary}

The images of the new TNOs discovered by CFEPS have been examined with a view to detect the presence of binaries. 
Both the methods of PSF subtraction and a newly developed scheme of 2D Gaussian profile fitting have been used. We only examine the objects brighter than $g \sim 23$, because both two methods are SNR sensitive.
Of the 63 new TNOs (31 of 63 are classical belt objects, 20 are the resonant objects) examined, two, L4k12 and L5c02 have been identified to be wide binaries and L5c02 have been confirmed. Both of them are classical TNOs.

To estimate the binary fraction in different dynamical classes, especially in cold and hot components of classical Kuiper belt, it is important to realize that an $i<5$ degree orbit could be drawn from either a low-inclination or higher-inclination component when both are modeled as gaussians.  We used the CFEPS survey simulator \citep{Kav09} to determine that for $g\lesssim23$ about 7\% of the $i<5^\circ$ detections are actually from the hot component, and thus it is problematic to label a particular TNO as 'cold' just because it has $i<5$ degrees.  Conversely, if L4k12 does turn out to be binary, with heliocentric $i\simeq35^\circ$ the probability it is drawn from the narrow inclination 'cold' population is essentially negligible.
In our 31 classical belt objects sample, 23 of 31 are $i<5^\circ$. Therefore we expect that $\simeq21$ are cold classical belt objects and 10 are hot components.  For the 95\% confidence interval, based on the Poisson uncertainty, we have constrained the binary rate to between 1.5\% and 20\% for the 'cold belt' and between 3 and 43\% for the hot belt. Also, the 95\% confidence interval for 0 detections is an upper limit of 3 sources, we know the binary fraction among the 'resonant' sources is below 3/20, or 15\%.

This fraction is substantially lower than that determined from high-resolution HST imaging \citep[see][]{Nol08b} but is consistent with the fraction of wider separation systems reported in \citet{Ker06}.
We note that L4k12 is has the highest heliocentric orbital inclination among all currently known binary sources.

The present numerical algorithm will be a promising approach in the search of TNBs in large surveys like Pan-STARRS which is expected to multiply the number of detected TNOs many times. 
A systematic search as described in the present work will shed new light on the orbital distribution of TNBs and their origin.   

Additional high quality tracking observations of L4k12 from GEMINI (A.H. Parker, private communications) have not resolved (seeing FWHM $\sim 0.5\arcsec$) this source as binary. Given the lack of a followup observations of our Subaru confirmation data we caution that L4k12 may actually be a tight binary (separation $<0.5\arcsec$) discovered in the CFEPS images during its apocentre passage.  We are continuing to monitor this source in an attempt to fully resolve the nature of this putative TNB.

\acknowledgments
This work was supported in part by NSC Grant: NSC 97-2112-M-008-011-MY3 and Ministry of Education under the Aim for Top University Program NCU.

\begin{deluxetable}{rrrrrrrr}
\tablecolumns{8}
\tablewidth{0pc}
\tablecaption{Heliocentric orbital data and system magnitudes}
\tablehead{\colhead{Object} & \colhead{a(AU)}   & \colhead{e}    & \colhead{i(degree)} & \colhead{Dynamical Class}    & \colhead{Magnitude}  & \colhead{Mutal Period(day)*}}
\startdata
L4k12 & 40.77 & 0.117 & 35.23 & Classical & 23.0 (r') & 320 \\
L5c02 & 45.74 & 0.035 & 1.791 & Classical & 22.8 (r') & 314\\
\enddata
\tablecomments{*Assume that semi-major axis equal to separation of two components with  $2g/cm^{3}$ bulk density. }
\label{tab:log}
\end{deluxetable}

\begin{deluxetable}{rrrrr}
\tablecolumns{8}
\tablewidth{0pc}
\tablecaption{Observation summary}
\tablehead{\colhead{Telescope} & \colhead{UT Date}   & \colhead{$\Delta$ mag}    & \colhead{Separation('')} & \colhead{PA(degree)*}}
\startdata
\multicolumn{5}{l}{L4k12}\\
 CFHT   & 2005 May 14.4 & \nodata & 0.4 & 270  \\
 Subaru & 2009 Apr 23.6 & 0.71& 0.79 & 270 \\
\multicolumn{5}{l}{L5c02}\\
CFHT  & 2005 Feb 10.3 &  0.28 &   0.61    &28  \\
CFHT  & 2005 Apr 11.3 &  0.26 &  0.62     &   35\\
CFHT  & 2006 Feb 06.3 & 0.13  &   0.75    &  95 \\
WIYN  & 2008 May 03.2 &  0.32 & 0.67 & 207  \\
\enddata
\tablecomments{PA is angle east of north.}

\label{tab:follow}
\end{deluxetable}

\addcontentsline{toc}{chapter}{Bibliography}
\begin{thebibliography}{99}
\bibitem[Elliot et al. (2005)]{Ell05}
   Elliot, J.L., Kern, S.D., Clancy, K.B., Gulbis, A.A.S., Millis, R.L., Buie, M.W., Wasserman, L.H., Chiang, E.I., Jordan, A.B., Trilling, D.E., and Meech, K.J., 2005, \apj, 129, 1117-1162.
\bibitem[Gladman et al. (2008)]{Ga08}
  Gladman, B.J., Marsden, B.G. and VanLaerhoven, C.,  "The Solar System Beyond Neptune", 43-57, University of Arizona Press. 
\bibitem[Jewitt \& Luu (1992)]{Jew92}
   Jewitt, D.C. and Luu, J.X., 1992, Nature, 362, 730-732.
\bibitem[Jones et al. (2006)]{Jon06}
   Jones, R.L., Gladman, B., Petit, J.-M., Rousselot, P., Mousis, O., Kavelaars, J.J., Parker, J.Wm, Nicholson, P., Holman, M., Doressoundiram, A., Veillet, C., Scholl, H., and Mars, G., 2006, Icarus, 185, 508-522.
\bibitem[Kavelaars et al. (2009)]{Kav09}
   Kavelaars, J. J., Jones, R. L., Gladman, B. J., Petit, J.-M., Parker, Joel Wm., Van Laerhoven, C., Nicholson, P., Rousselot, P., Scholl, H., Mousis, O., Marsden, B., Benavidez, P., Bieryla, A., Campo Bagatin, A., Doressoundiram, A., Margot, J. L., Murray, I. and Veillet, C., 2009, \aj, 137, 4917-4935.
\bibitem[Kern and Elliot (2006)]{Ker06}
   Kern, S. D. and Elliot, J. L., 2006, \apj, 643, 57-60.
\bibitem[Millis et al. (2002)]{Mil02}
   Millis, R.L., Buie, M.W., Wasserman, L.H., Elliot, J.L., Kern, S.D., and Wagner, R.M., 2002, \aj, 123, 2083-2109.
\bibitem[Noll et al. (2008a)]{Nol08a}
   Noll, K.S., Grundy, W.M., Chiang, E.I., Margot, J.-L. and Kern, S.D., 2008, "The Solar System Beyond Neptune", 345-363, University of Arizona Press. 
\bibitem[Noll et al. (2008b)]{Nol08b}
   Noll, K.S., Grundy, W.M., Stephens, D.C., Levison, H.F. and Kern, S.D., 2008, Icarus, 194, 758-768.
\bibitem[Osip et al. (2008)]{Osi08}
   Osip, David J., Kern, S. D. and Elliot, J. L., 2003, EM\&P, v. 92, Issue 1, p. 409-421.
\bibitem[Petit et al. (2008)]{Pet08}
   Petit, J.-M., Kavelaars, J.J., Gladman, B.J., Margot, J.L., Nicholson, P.D., Jones, R.L., Parker, J.Wm., Ashby, M.L.N., Campo Bagatin, A., Benavidez, P., Coffey, J., Rousselot, P., Mousis, O. and Taylor, P.A., 2008, Science 322, 432.
\bibitem[Petit et al. (2010)]{Pet10}
   Petit, J.-M., Kavelaars, J.J., Gladman, B.J.,  Jones, R.L.,  Parker, J.Wm., VanLaerhoven, C., Nicholson, P., Mars, G., Nicholson, P.,  Mousis, O., Marsden, B.,  Bieryla, A., Murray, I., Ashby, M.L.N., Benavidez, P., Campo, A., Bagatin, Bernabeu, G., 2010, \aj, submitted
\bibitem[Stephens et al. (2006)]{Ste06}
   Stephens, D.C. and Noll, K.S. 2006, \aj, 131, 1142-1148.
\bibitem[Trujillo \& Brown (2003)]{Tru03}
   Trujillo, C. A. and Brown, M. E., EM\&P, 2003, v. 92, Issue 1, p. 99-112.
\bibitem[Veillet et al. (2002)]{Vei02}
   Veillet, C., Parker, J.W., Griffin, I., Marsden, B., Boressoundiram, A., Buie, M., Tholen, D.J., Connelley, M., Holman, M.J., 2002, Nature, 416, 711-713.
\end{thebibliography}
\end{document}